\documentclass[aps,prl,reprint,groupaddress]{revtex4-2}
\usepackage{dcolumn,graphicx,physics,braket,hyperref}
\usepackage{color}
\usepackage{hyperref}

\newcommand{\bk}{\mathbf{k}}

\begin{document}

\title{Giant proximity exchange and flat Chern band in 2D magnet-semiconductor heterostructures}

\author{Nisarga Paul, Yang Zhang, Liang Fu}

\affiliation{Department of Physics, Massachusetts Institute of Technology, Cambridge, MA,
USA}


%

\begin{abstract}
Van der Waals (vdW) heterostructures formed by two-dimensional magnets and semiconductors have provided a fertile ground for fundamental science and spintronics. We present first-principles calculations finding a proximity exchange splitting of 14 meV (equivalent to an effective Zeeman field of 120 T) in the vdW magnet-semiconductor heterostructure MoS$_2$/CrBr$_3$, leading to a 2D spin-polarized half-metal with carrier densities ranging up to $10^{13}$ cm$^{-2}$. We consequently explore the effect of large exchange coupling on the electronic bandstructure when the magnetic layer hosts chiral spin textures such as skyrmions. A flat Chern band is found at a ``magic" value of magnetization $\overline{m} \sim 0.2$ for Schr\"odinger electrons, and it generally occurs for Dirac electrons. The magnetic proximity induced anomalous Hall effect enables transport-based detection of chiral spin textures, and flat Chern bands provide an avenue for engineering various strongly correlated states.
\end{abstract}
\maketitle

\section{Introduction}

Flat band materials have recently emerged as an area of intensive study in condensed matter physics \cite{Andrei2020Dec,Heikkila2011Oct,Tang2014Dec}. Due to the dominance of many-body effects over the quenched kinetic energy, such systems provide avenues for realizing strongly correlated electronic states such as generalized Wigner crystals, magnetic orders, and superconductivity. At partially filling of a topological flat band, interaction effects can further lead to quantum anomalous Hall states \cite{Serlin2020Feb,Sharpe2019Aug,Chen2020Mar,Li2021Jul} as  proposed in twisted bilayer graphene (TBG) \cite{PhysRevResearch.1.033126,PhysRevLett.124.097601} and transition metal dichalcogenide (TMD) moir\'e heterostructures \cite{PhysRevLett.122.086402,Devakul2021Jun,Zhang2021Sep}. A common element of moir\'e materials is the presence of a spatially varying periodic modulation, e.g. of interlayer tunneling strength, which generates narrow or flat minibands.

\par 

Such a modulation is also present in magnets hosting periodic spin textures, which couple to electrons as a Zeeman splitting with spatially varying orientation. Noncoplanar or chiral spin textures, including skyrmions and canted spirals or vortices, are characterized by their nonzero spin chirality defined as $\vec S_i \cdot (\vec S_j \times \vec S_k)$. Chiral spin textures have been the focus of recent interest due to the observation of skyrmions in a host of compounds and a large anomalous Hall effect in noncoplanar magnets \cite{Muhlbauer2009,Yu2012,Nagaosa2013,Buttner2018,machida2010time,nakatsuji2015large,nayak2016large,Nagaosa2013,Liu2022Jul,lado}. Electrons coupled to skyrmions experience an emergent magnetic field which manifests in the topological Hall effect \cite{Bruno2004Aug,Neubauer2009May}. The presence of electron minibands due to a superlattice modulation is common to both structural moir\'es and periodic spin textures \cite{Paul2021Aug,PhysRevResearch.3.033156,Divic2021Mar,Chen2021Sep}, but the close analogy between the two types of systems has not yet been explored.


\par 
\begin{figure}
    \centering
    \includegraphics[width=\linewidth]{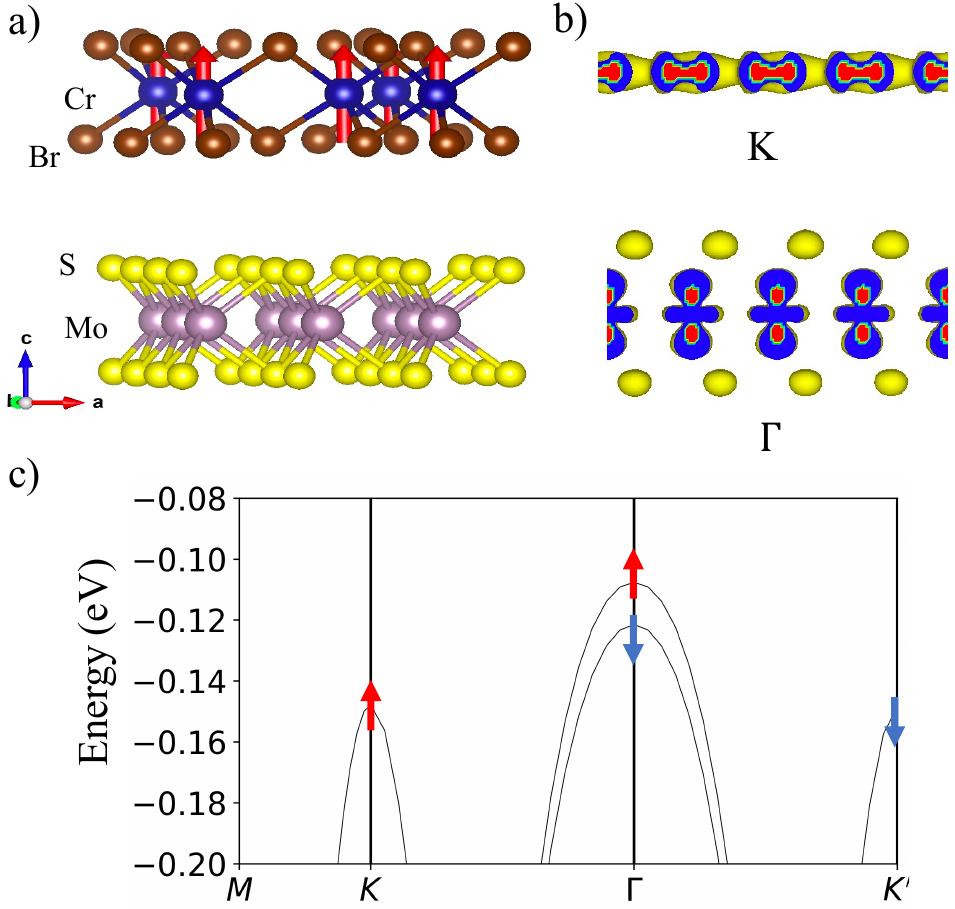}
    \caption{\textbf{The large exchange splitting in MoS$_2$/CrBr$_3$ as a route to flat Chern bands.} (a) Lattice structure and magnetic order of MoS$_2$/CrBr$_3$ heterostructures. (b) Kohn-Sham wavefucntion for $K$ valley and $\Gamma$ valley maximum. The $K$ valley wavefunction is localized in the Mo layer, while the $\Gamma$ valley wavefunction is spread out in the z direction. (c) Band structure of the fully relaxed MoS$_2$/CrBr$_3$ heterostructure with a 0.9 meV valley splitting at $K^{\prime}$/$K$ and 14 meV spin splitting at $\Gamma$.}
    \label{fig:1}
\end{figure}

A natural question is whether chiral spin textures can provide a route to realizing \textit{flat} bands. Indeed, the spatial profile of spin textures can be tuned by externally controllable parameters, such as magnetization and period, which may function similarly to twist angle or lattice mismatch in TBG or TMDs in optimizing band flatness. Moreover, the emergent magnetic field may generate mini-bands with Chern number \cite{Hamamoto2015}, furthering the analogy with structural moir\'es. A possible hurdle is that the exchange coupling between localized spins and itinerant electrons must be large for emergent field to have a strong effect on the electronic band structure. 
In reality, however, most skyrmion materials are high density metals in which the exchange splitting is small compared to the Fermi energy.  
 
 \par 

In this work, we propose a new platform for a topological flat band in 2D magnet/ TMD semiconductor heterostructures, in which the magnetic exchange coupling introduces an emergent Zeeman field acting on a low density of carriers in the semiconductor. Our study is motivated by the recent theoretical  \cite{Tong2018Nov, Hejazi2020May,Lu2020Sep,Augustin2021Jan} and experimental  works \cite{Xu2021Mar} suggesting the existence of skyrmion phases and periodic chiral spin textures arising from a spatially varying interlayer coupling in magnetic moir\'es such as twisted bilayers of CrI$_3$ or CrBr$_3$.  
Our two main results are as follows. First, we predict from first-principles calculations a giant exchange splitting of $14$ meV in MoS$_2$/CrBr$_3$, which far exceeds the values found so far in 2D TMD/magnet heterostructures. 
As a result of this large exchange splitting, for a wide range of hole densities up to $0.83\times 10^{13}$cm$^{-2}$ only one spin-polarized band is occupied, leading to a 2D half metal.  
Second, we use a continuum model approach to study the effect of the emergent field from magnetic skyrmion crystals on the low-energy electronic structure of proximitized 2D semiconductors. We find an almost completely flat Chern band for Schr\"odinger electrons (as in MoS$_2$) at a ``magic" value of magnetization $\overline{m}=0.2$, while for Dirac electrons (as in graphene), remarkably, a flat Chern band always occurs. 

\par 



\section{Results}

\textit{Giant exchange splitting in MoS$_2$/CrBr$_3$.---} Our proposal to realize flat Chern bands requires a strong interfacial exchange coupling in 2D magnet-TMD semiconductor heterostructures. In monolayer TMDs, strong Ising spin-orbit coupling and inversion symmetry breaking gives rise to opposite spin states in the $K^{\prime}$ and $K$ valleys. 
Previous studies have demonstrated $K^{\prime}$/$K$ spin-valley splitting in TMDs proximity coupled to ferromagnetic semiconductors. However, such splitting is generally small, e.g.,  
3.5 meV in WSe$_2$/CrI$_3$ \cite{zhong2017van,Lyons2020Nov}. 

We performed density functional calculations using the generalized gradient approximation \cite{perdew1996generalized} with SCAN+rVV10 van der Waals density functional \cite{peng2016versatile}, as implemented in the Vienna Ab initio Simulation Package \cite{kresse1996efficiency}.
By scanning through various 2D TMD-magnet heterostructures, we find a giant exchange splitting in MoS$_2$/CrBr$_3$, where monolayer MoS$_2$ is a TMD semiconductor, and CrBr$_3$ is an insulating vdW magnet. Magnetic orders and lattice constants are taken from the Computational 2D Materials Database (C2DB) \cite{haastrup2018computational}. The 2x2 supercell of MoS$_2$ with lattice constant 6.38 \AA\, is chosen to match the lattice constant of CrBr$_3$ (6.44 \AA) with less than 1\% strain. Similar to previous works, we find a small $K^{\prime}$/$K$ valley splitting on the valence band side of 0.9 meV for MoS$_2$/CrBr$_3$.

In monolayer MoS$_2$, the $\Gamma$ valley is very close in energy to the $K$ valley. In MoS$_2$/CrBr$_3$, we find that the $\Gamma$ valley of MoS$_2$ becomes the valence band maximum after taking into account the relaxed heterostructure with layer spacing 6.76 \AA \, (defined as the distance between the center of MoS$_2$ and CrBr$_3$). We observe a giant spin splitting of up to 14 meV in the $\Gamma$ valley as shown in Fig. \ref{fig:1}. Assuming a Land\'e $g$-factor equal to 2, this is an effective Zeeman field of 120 T. (For MoS$_2$ on twisted bilayer CrBr$_3$, which is relevant for skyrmion physics, we find a similarly giant spin splitting of 17 meV; see the Supplemental Materials). The large difference in the exchange splitting in the $K$ and $\Gamma$ valleys can be understood from the spatial distribution of wavefunctions. At the $K$ valley, due to the large in-plane momentum, electron wavefunctions are strongly localized within the vdW layer. However, at the $\Gamma$ valley, due to the zero in-plane momentum, the wavefunctions are more spread out in out-of-plane direction, which enhances coupling to the magnetic layer. 
The large exchange splitting in this specific material, which is the first main result of this work, provides efficient spin control of the $\Gamma$ valley bands. 
\par 


\textit{Emergent gauge field at large exchange coupling.---} In the following, we explore the consequences of a large exchange coupling on the electronic bandstructure. This is motivated by both the giant exchange splitting found in the previous section and theoretical and experimental results suggesting that twisted bilayers of vdW magnets may host noncoplanar or skyrmionic phases \cite{Tong2018Nov, Hejazi2020May,Lu2020Sep,Augustin2021Jan,Xu2021Mar}. To this end, we turn to a completely general Hamiltonian describing electrons coupled to a spin texture via a large exchange coupling. 
It is important to note that in low density systems such as TMD semiconductors or graphene, 
the physics of itinerant electrons coupled to periodic spin textures 
is universally described by a continuum theory, taking the form


\begin{subequations}
\begin{align}\label{eq:HS}
H\textsubscript{S} &= \frac{p^2}{2m} + J \vec\sigma \cdot \vec S(r)\\
\label{eq:HD}
H\textsubscript{D} &= v\vec p \cdot \vec \tau + J \vec \sigma \cdot \vec S(r)
\end{align}
\end{subequations}

\noindent for the case of Schr\"odinger and Dirac electrons with mass $m$ and velocity $v$, respectively. The Pauli $\vec \sigma$ matrices act on electron spin, while the Pauli $\vec \tau$ matrices act on a psuedospin degree of freedom, e.g., sublattice in the case of graphene. Eq. \ref{eq:HS} describes TMD semiconductors proximity-coupled to a magnetic layer, while Eq. \ref{eq:HD} describes Dirac materials such as graphene and twisted bilayer graphene.\par

In the following, we will consider smoothly varying periodic spin textures $\vec S(r)$ with magnetic wavelength $\xi$ and unit norm $|\vec S|=1$. An SU(2) gauge transformation $\mathbf{U}(r)$ allows us to instead consider a uniformly polarized spin texture provided we use the gauge-transformed momentum $\vec p \mathbf{1}-e \vec{\mathbf{A}}$, where 

\begin{equation}
        \vec{\mathbf{A}} = \frac{i\hbar}{e} \mathbf{U}^\dagger \vec \nabla\mathbf{ U}.
\end{equation}

\par 
\noindent The large exchange coupling limits are, respectively,
\begin{equation}\label{eq:lim}
    Jm\xi^2 \gg 1\quad \text{ and } \quad J\xi/v \gg 1.
\end{equation}
In this limit the ground state manifold consists of electrons with spins locally anti-aligned with the spin texture. The effective Hamiltonian describing the locally spin-polarized Schr\"odinger electrons has been derived before \cite{Bruno2004Aug}:

\begin{equation}
    {\tilde H}\textsubscript{S} = \frac{1}{2m} (\vec p-e \vec A)^2 + \frac{e^2}{8m}(\partial_i\vec S)^2
\end{equation}
(summing $i=x,y$). The second term, originating from off-diagonal terms of $\vec{\mathbf{A}}$, is a scalar potential that accounts for the reduction of electron hopping by the spin gradient 
and $\vec A= [\vec{\mathbf{A}}]_{\downarrow \downarrow}$ is an emergent $U(1)$ gauge field. The corresponding emergent field strength

\begin{equation}
    B^e =\frac{\hbar}{2e}\vec S \cdot (\partial_x \vec S \times \partial_y \vec S)
\end{equation}
is proportional to the spin chirality. A skyrmion can be defined as a texture with flux $h/e$. It follows from this equation that purely in-plane spin textures have vanishing emergent field. Moreover, purely in-plane textures are symmetric under the combination of time reversal and a spin-space $C_2$ rotation, while the Chern number is odd under this combination, implying such textures do not yield topological bands. Accordingly, we will consider chiral magnets with out-of-plane spin textures. 
\begin{figure}
    \centering
    \includegraphics[width=\linewidth]{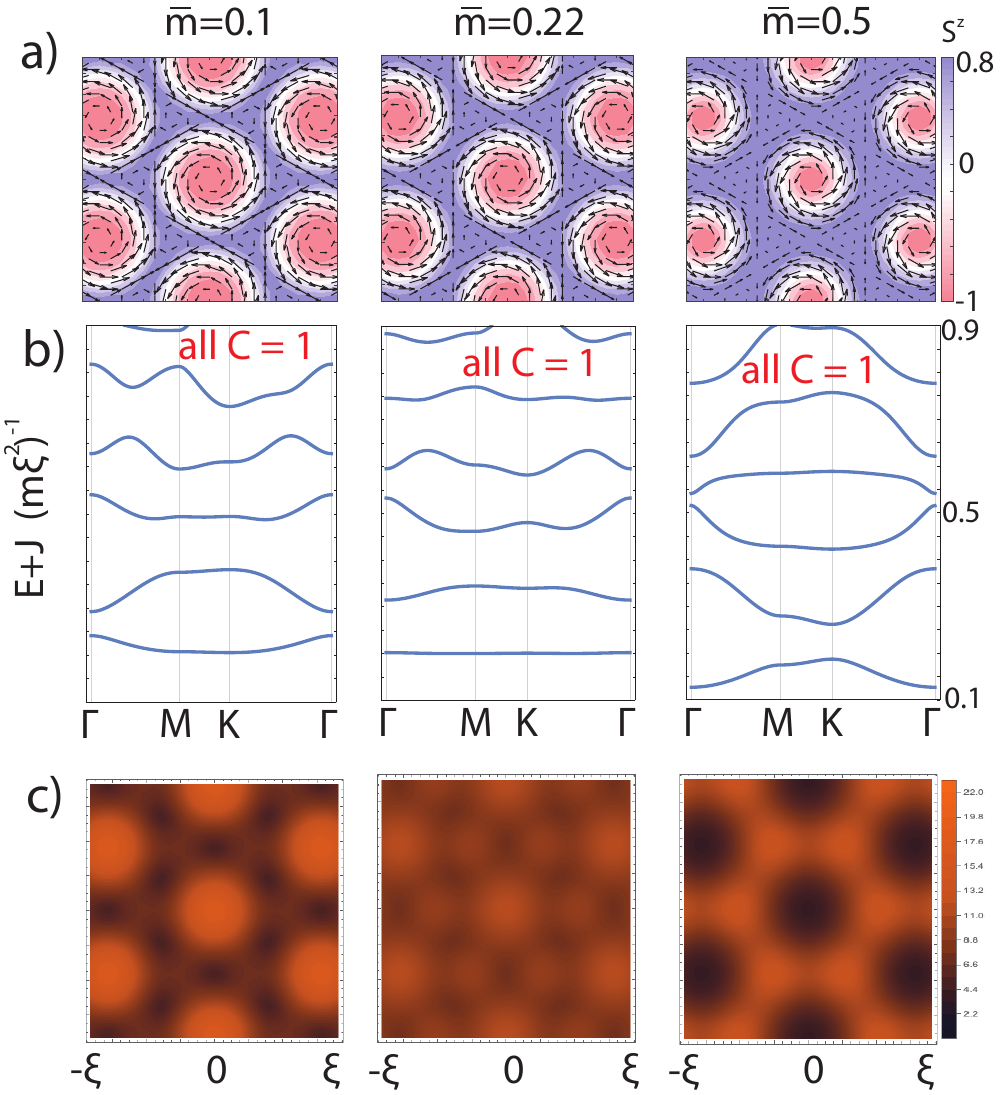}
    \caption{\textbf{Flat Chern band from a skyrmion crystal: Schr\"odinger electrons.} (a) Skyrmion texture based on generic ansatz Eq. \eqref{eq:skx} with different magnetizations and $\alpha=1$. $(S^x,S^y)$ is plotted as a vector field with $S^z$ in color, and $\overline{m}$ is the average of $S^z$. (b) Bandstructures in the presence of the textures in (a), following the $\Gamma MK\Gamma$ path in the hexagonal mini BZ. $H_S$ was diagonalized using a plane-wave cutoff $N=10$, taking $J=100,m=1$. 
    A flat first band is observed when $\overline{m}=0.22$. All bands carry Chern number $C=1$. (c) Local density of states summed over the first band (sampling $4^2$ states), spatially aligned with (a). }
    \label{fig:2}
\end{figure}

We will also be interested in Dirac electrons at large exchange coupling, for which the effective Hamiltonian takes the form


\begin{equation}\label{eq:Hdeff}
    {\tilde H}\textsubscript{D} = v(\vec p-e \vec A)\cdot \vec \tau.
\end{equation}
It is notable that Dirac electrons only feel a periodic gauge field in contrast with the previously studied Schr\"odinger case involving both the gauge field and a scalar potential. \par 

The origin of the emergent gauge field is the local U(1) rotational freedom of the electron spin about the axis of the local spin.
In analogy with conventional Landau levels, the emergent field $B^e$ generally endows the minibands with Chern number when the average is nonzero. 

\par 
In the following, we investigate the flatness of the induced Chern bands in the large exchange coupling limit. How realistic is the large exchange coupling limit for the systems we have in mind? Our primary motivation is MoS$_2$/ CrBr$_3$, for which $J\approx $ 7 meV and $m^* \approx 1.1 m_e$.  Moir\'e magnets can exhibit periods $\xi$ of $10- 100$ nm \cite{Tong2018Nov,Akram2021Aug,Song2021Nov}. Altogether (restoring $\hbar$), we have $
    Jm\xi^2/\hbar^2 \sim 10 - 10^3$ for this system, i.e. we are \textit{deep into} the large exchange limit for this range of $\xi$, though theoretical estimates for skyrmion sizes in CrBr$_3$ are only approximate. \par 
 When $\xi$ is large, the kinetic energy scale $\hbar^2/m\xi^2$ is small and disorder may have a pronounced effect. However, all the low-lying bands are topological ($C=1$) bands, so a significant anomalous Hall effect is still expected \cite{Hamamoto2015}.
 
 For monolayer graphene, assuming $J \sim 1$ meV and $v = 2.5 \cdot 10^6$ m/s, $J\xi/\hbar v \sim 0.4$. However, the velocity can be sharply decreased in twisted bilayer graphene, and $\xi$ increased in a moir\'e magnet, so the large exchange limit can be achieved for Dirac materials as well. The large exchange limit is closely related to the adiabatic limit which has been frequently studied in the literature \cite{Denisov2016Jul,Nakazawa2018Feb,Denisov2017Dec}. A key difference is that our criterion makes no reference to electron density or scattering time, since we are working in the limit of pristine samples and low densities on the order of one charge per SkX unit cell.


\textit{Flat bands in skyrmion crystals.---} We present numerical results on the bandstructure of Schr\"odinger and Dirac electrons in the presence of generic skyrmion crystals using the full Hamiltonians in Eqs. \eqref{eq:HS} and \eqref{eq:HD}, i.e. without any approximations. $H_S$ and $H_D$ are explicitly diagonalized in a plane wave basis with a large wavenumber cutoff $N$ (see Supplemental Material). 


As a model for a generic SkX, we introduce the simple three-parameter ansatz $\vec S= \vec N_r / |\vec N_r|$ with 

\begin{equation}\label{eq:skx}
    \vec N_r =\frac{1}{\sqrt{2}}\sum_{j=1}^6 e^{i q_j\cdot r}\hat e_j + \mu \hat z
\end{equation}
 where $\vec q_j = \xi(\cos \theta_j,\sin\theta_j)$ and $\hat e_j=(i\alpha\sin\theta_j,-i\alpha\cos\theta_j,-1)/\sqrt{2}$ and the angles satisfy $\theta_2=\theta_1+2\pi/3,\theta_3=\theta_1+4\pi/3$, and $\theta_{j+3}=\theta_j+\pi$. Eq. \eqref{eq:skx} represents a normalized sum of three helical spirals forming a  triangular SkX, as plotted in Fig. \ref{fig:2}a. This ansatz is widely adopted in studies of chiral magnets and magnetic skyrmion crystals. It qualitatively reproduces the observed real-space images of skyrmion crystals \cite{Karube2017Dec,Park2011May,Tokura2020Nov,Lin2016Feb,Shimizu2021May,Yu2010Jun,Tonomura2012Mar}. A different choice of relative phases results in N\'eel-like skyrmions but doesn't alter bandstructures. The parameters $\alpha, \xi,$ and $\mu$ control coplanarity, wavelength, and out-of-plane bias, respectively. 

In Fig. \ref{fig:2} we plot bandstructures and local density of states (LDOS) for Schr\"odinger electrons in the presence of our chosen SkX. We set $\alpha=1$ and vary the magnetization $\overline m$ (i.e. the unit cell average of $S^z$), which is monotonic with $\mu$ ($\xi$ can be scaled out of the problem). We can observe the flattening of lower bands and a near perfectly flat, well-separated first band as we tune the magnetization past $\overline{m}=0.2$, the second main result of this work. All bands shown carry Chern number $C=1$ and spins are anti-aligned with the skyrmion texture.

\begin{figure}
    \centering
    \includegraphics[width=\linewidth]{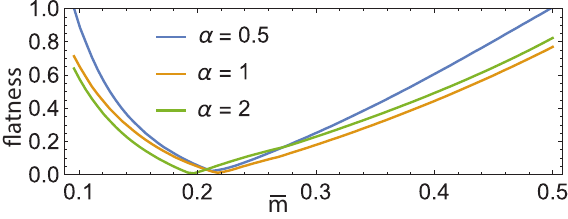}
    \caption{\textbf{Flatness at a special magnetization.} First band flatness, defined as the ratio of bandwidth to bandgap, of $H_S$ in the presence of a generic skyrmion crystal (defined in Eq. \eqref{eq:skx}) with varying magnetization $\overline{m}$ and coplanarity $\alpha$. Near perfect flatness occurs around $\overline{m}=0.2$. Flatness was calculated by solving $H_S$ with plane-wave cutoff $N=10$ and $J=100$ over $10^2$ points in the mini BZ.}
    \label{fig:3}
\end{figure}

In Fig. \ref{fig:3} we systematically plot band flatness, defined as the ratio of bandwidth to bandgap for the first miniband, with varying $\alpha$ and $\overline{m}$. Complete flatness is an ideal pathway for strongly correlated matter since the kinetic energy is strongly quenched and band separation remains large. We plot the branch $\overline{m}>0$ which corresponds to the typical case of separated skyrmion bubbles. The plot indicates a quite robust ``magic" value of magnetization near $\overline{m}=0.2$ where flatness goes to zero. 


\begin{figure}
    \centering
    \includegraphics[width=\linewidth]{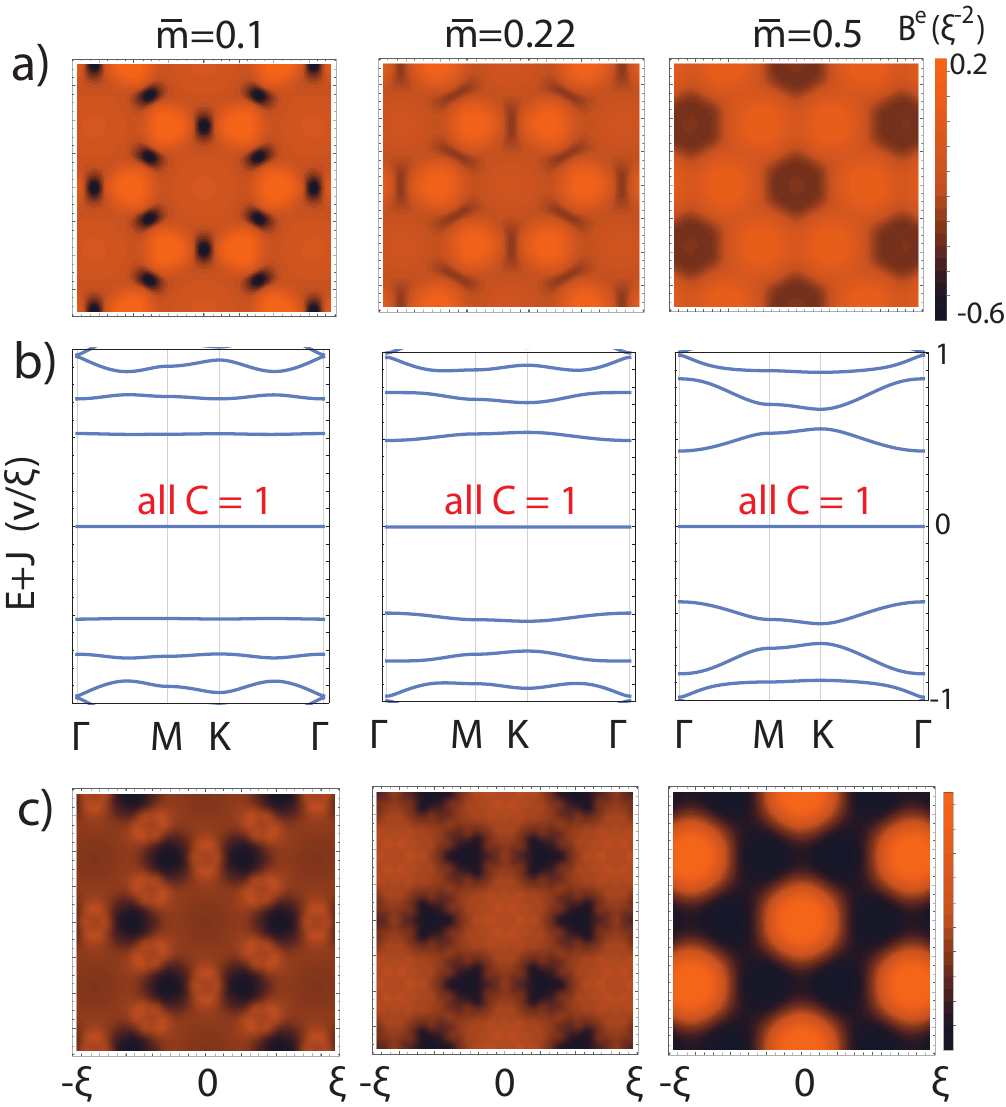}
    \caption{\textbf{Flat Chern band from a skyrmion crystal: Dirac electrons.} (a) Emergent magnetic fields corresponding to (and spatially aligned with) the SkX shown in Fig. \ref{fig:2}a. Dirac electrons in the large exchange coupling limit move in this emergent field. (b) Bandstructures analogous to Fig. \ref{fig:2}b for Dirac electrons. A flat band is pinned at $E=-J$ whenever there is net emergent flux. All bands carry Chern number $C=1$.} (c) Local density of states summed over the $E=-J$ band (sampling $150$ states), spatially aligned with (a).
    \label{fig:4}
\end{figure}

Next we turn to the case of Dirac electrons under periodic Zeeman field, which may be realized in graphene or twisted bilayer graphene in proximity with a 2D magnet. Remarkably, we observe a flat band for \textit{arbitrary} skyrmion textures in Fig. \ref{fig:4}. This can be understood from the effective Hamiltonian in the large 
$J$ limit. 

\begin{equation}
    {\tilde H}\textsubscript{D} = v\begin{pmatrix}
    0 & \Pi\\
    \overline{\Pi} & 0
    \end{pmatrix}
\end{equation}
where $\overline{\Pi}=(\Pi_x+i\Pi_y)/2$ and $\vec \Pi = \vec p-e\vec A$. 
It is known that the Dirac equation in the presence of a spatially varying magnetic field 
admits an extensive set of zero-mode solutions \cite{Aharonov1979Jun}. The number of zero modes is equal to the total number of flux quanta through the system (which in our case is given by the total topological charge of the skyrmion crystal), regardless of the spatial profile of the magnetic field. 
This remarkable property follows from the Atiyah-Singer index theorem. 
In the limit of a uniform field, these zero modes reduce to the well-known $n=0$ Landau level.

In Fig. \ref{fig:4}a we plot the periodic emergent magnetic fields arising from the skyrmion crystal  and in Fig. \ref{fig:4}c we plot the LDOS of the flat band of Dirac electrons under magnetic proximity effect. The LDOS shows significant spatial variation, which can be detected by scanning tunneling microscopy.   

\par 

In real magnetic moir\'e systems, there may be magnetic disorder or a lack of strict periodicity in the magnetic domains \cite{Song2021Nov}. A general expectation is that the lack of a strict periodicity would broaden the band structure, which were all calculated for pristine samples. Still, we may expect a DOS enhancement even with disorder. Moreover, Dirac flat band is actually immune to magnetic disorder, and an extensive set of zero modes would exist even in the absence of periodicity. For the flat band, the only requirement is a net skyrmion density; one Dirac zero modes will bind to each skyrmion, and we expect the strongly correlated physics associated with a DOS enhancement to survive disorder in this case.

\textit{Small exchange coupling and anomalous Hall effect.---}
While the large exchange coupling regime is realistic for MoS$_2$/CrBr$_3$, other systems may have constraints on magnetic wavelength or intrinsic exchange coupling not conducive to this limit. Additionally, many chiral magnets host noncoplanar states which do not enclose a net spin chirality. Examples of such states include the canted vortex lattice, multiple-Q conical spiral, multilayers of single-Q helices, and meron-antimeron lattice \cite{Lu2020Sep,Augustin2021Jan}. \par

\begin{figure}[t]
    \centering
    \includegraphics[width=1.0\linewidth]{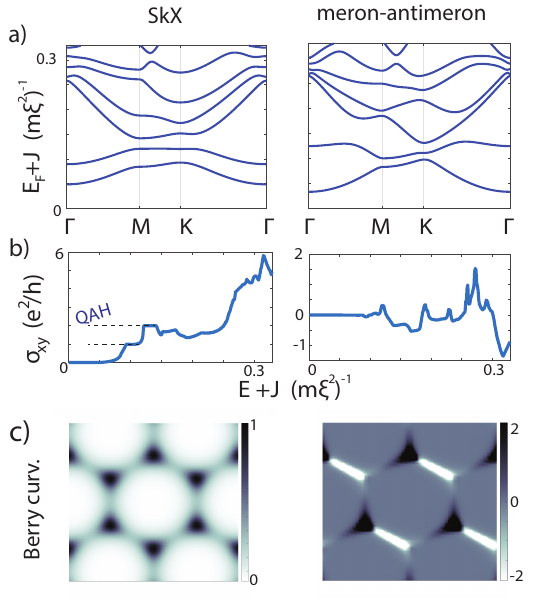}
    \caption{\textbf{Band topology at weak coupling.} (a) Weak-coupling bandstructures for the SkX ($\overline{m}=0.1,\alpha=1$) and the  meron-antimeron lattice. Each band in the former has Chern number 1, and the lowest bands in the latter have Chern numbers [0,0,1,-1,1]. (b) Hall conductances. The SkX exhibits a quantized anomalous Hall effect from the lowest bands, and the meron-antimeron lattice  has a nonzero anomalous Hall response despite its vanishing net spin chirality. (c) Similarly for an SkX (left) and meron-antimeron lattice (right) at weak exchange coupling, corresponding to the lowest bands of (a)}
    \label{fig:5}
\end{figure}

In the large exchange coupling regime, we found well-separated flat Chern bands (which imply a quantized anomalous Hall effect) in the presence of skyrmion-like textures. If we drop these two prerequisites and instead consider small exchange coupling and general noncoplanar textures (i.e. not necessarily enclosing a net spin chirality), we find numerically that an anomalous Hall conductance is still generically present. We illustrate this point with an SkX and a meron-antimeron lattice at small exchange coupling (Fig. \ref{fig:5}a), which takes the form

\begin{equation}
    \vec S_r = (\sqrt{1-(S^z)^2}\sin\theta_r,\sqrt{1-(S^z)^2} \cos\theta_r, S^z)
\end{equation}
with $\theta_r$ the angle relative to the unit cell center and $S^z$ is a smooth periodic function which equals unity at the unit cell centers. We chose $S^z(r)= \frac13 \sum_i \cos^2(q_i\cdot r/2)$ where $q_1=b_1,q_2=b_2, q_3=-b_1-b_2$ and $b_{1,2}$ are reciprocal lattice vectors. This texture has alternating regions of $\pm 1/2$ emergent flux, justifying the name meron-antimeron lattice (and has only three-fold symmetry). Because the net spin chirality vanishes, and na\"ively we may expect no momentum-space topology. However, we observe miniband Chern numbers and a corresponding anomalous Hall effect (Fig. \ref{fig:5}b). Moreover, we have produced plots of the Berry curvatures of the SkX and meron-antimeron lattice at weak exchange couplings (Fig. \ref{fig:5}c).

\par

\begin{figure}[t]
    \centering
    \includegraphics[width=\linewidth]{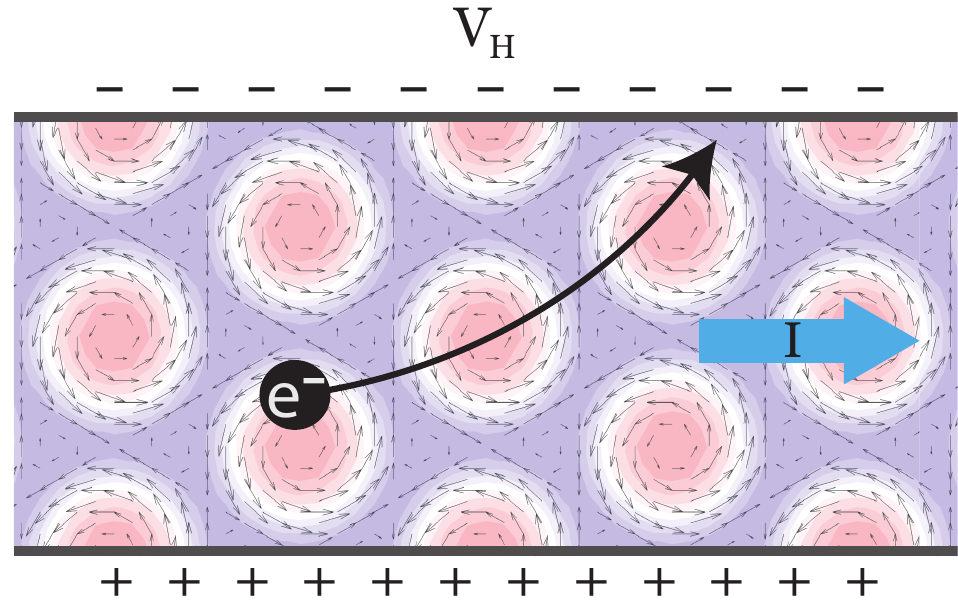}
    \caption{\textbf{AHE from real-space topology.} An anomalous Hall effect is generically expected in semiconductor / chiral magnet heterostructures with noncoplanar textures even at small coupling and without net spin chirality.}
    \label{fig:hall}
\end{figure}

The observation of marked Hall transport features suggests that proximity-coupled electrons can serve as a reliable sensor for noncoplanar textures in magnetic materials (Fig. \ref{fig:hall}). A transport-based sensor would be advantageous because it can be done on tabletop while bulk probes such as Lorentz TEM and neutron diffraction have a low sensitivity for 2D devices. We also note that in the presence of a point-group symmetry, the leading Fourier modes of the spin texture may be inferred from the miniband Chern numbers. This provides a useful direct mapping between real-space textures and transport properties. For instance, in the presence of six-fold symmetry the lowest band's Chern number satisfies

\begin{equation}
C \bmod 6 = \begin{cases}
1 & S_0^z>0, S_{\textbf{G}}^z>0\\
2 \text{ or } 0 & S_0^z>0, S_{\textbf{G}}^z<0\\
-2 \text{ or } 0 & S_0^z<0, S_{\textbf{G}}^z>0\\
-1 & S_0^z <0, S_{\textbf{G}}^z <0\\
\end{cases}
\end{equation}
 assuming $J>0$. For $J<0$ the Chern number changes sign mod 6. Here $\vec S_0, \vec S_{\textbf{G}}$ are the zeroth and first Fourier components of $\vec S(r)$, so in particular $S_0^z=\overline{m}$. Here $\textbf{G}$ is one of the primitive reciprocal lattice vectors ($\vec S_{\textbf{G}}$ doesn't depend on which one, by six-fold symmetry). For fixed $S_{\textbf{G}}^z$, it follows from the above that the first band undergoes a topological transition as the magnetization changes sign, which can be traced to the gap closing at $\Gamma$. Likewise, for fixed magnetization there is a topological transition as $S_{\textbf{G}}^z$ changes sign. We leave the derivation to the Supplemental Material. 

\par

\section{Discussion}

 We investigated the emergence of flat bands in semiconductors proximity-coupled to spin textures in the limit of large exchange coupling and emerged with a concrete prediction: that the TMD/ vdW magnet heterostructure MoS$_2$ proximity coupled to twisted bilayer CrBr$_3$ should exhibit a flat Chern band at magnetization $\overline{m} \sim 0.2$. Our study was motivated by the giant exchange of 14 meV (corresponding to a large exchange parameter $Jm \xi^2/\hbar^2 \sim 10^3$) which we found using first-principles calculations in MoS$_2$/CrBr$_3$, the latter of which is proposed to host skyrmions in a moir\'e \cite{Hejazi2020May,Tong2018Nov}. For Schr\"odinger electrons coupled to a generic SkX we find a flat Chern band emerges at out-of-plane magnetization $\overline{m}=0.2$ while a flat Chern band is generically present for Dirac electrons. A large exchange coupling and long electron mean free path (for sharply defined minibands) are prerequisites for this route to flat bands. 
These continuum results are universal at low densities, only making use of effective mass or Dirac velocity in the band structure. We show explicitly that anomalous Hall effect is generically present in TMD semiconductors or graphene proximitized by noncoplanar magnets. 
Our results also suggest the possibility of fractional Chern insulators (FCIs) in 2D magnet-semiconductor or graphene heterostructures. 
\par 

\section{Materials and Methods}

\textit{DFT: } We performed the density functional calculations using generalized gradient approximation \cite{perdew1996generalized} with SCAN+rVV10 van der Waals density functional \cite{peng2016versatile}, as implemented in the Vienna Ab initio Simulation Package \cite{kresse1996efficiency}. Pseudopotentials are used to describe the electron-ion interactions.  We first constructed zero degree aligned 2x2 MoS$_2$/CrBr$_3$ heterobilayer with vacuum spacing larger than 20 A to avoid artificial interaction between the periodic images along the $z$ direction. Dipole corrections were added to the local potential in order to correct the errors introduced by the periodic boundary conditions in out of plane direction. The structure relaxation was performed with force on each atom less than 0.001 eV/A. We used a dense kmesh sampling up to $12\times 12$ for structure relaxation and self-consistent calculation. In the Supplementary Materials we detail DFT calculations with the addition of a Hubbard $U$ and for MoS$_2$ on bilayer CrBr$_3$, and find that our main conclusions are unchanged. \par 
\textit{Bandstructure: } The plane wave method was used for Fig.s \ref{fig:2}, \ref{fig:3}, and \ref{fig:4} while for \ref{fig:5} a square lattice nearest neighbor tight-binding Hamiltonian with parabolic low-energy dispersion was used. In the plane wave approximation (which we detail in the Supplementary Materials) we kept the $21^2$ lowest-lying Fourier modes of the skyrmion texture and electronic wavefunctions.

\textit{Acknowledgements:} We thank Joe Checkelsky, Takashi Kurumaji, Kin Fai Mak, Jie Shan, Riccardo Comin and Xiaodong Xu for helpful discussions, and Junkai Dong for collaboration on a related project. \par 
\textit{Funding:} This work is funded in part by the Air Force Office of Scientific
Research (AFOSR) under award FA9550-22-1-0432 and the Simons Foundation through a Simons Investigator Award. LF is partly supported by the David and Lucile Packard
Foundation. NP is partly supported by the U.S. Department of Energy, Office of Science, Basic Energy Sciences, under Award No. DE-SC0020149.\par 

\textit{Author contributions:} NP and LF contributed to the theoretical analysis and YZ contributed to the DFT analysis. NP wrote the initial version of the manuscript.\par

\textit{Competing interests:} All authors declare that they have no competing interests.\par

\textit{Data and Materials Availability:} All data needed to evaluate the conclusions in the paper are present in the paper and/or the Supplementary Materials.

\textit{Corresponding authors:} Email: liangfu@mit.edu (LF); npaul@mit.edu (NP)

\bibliography{flatchern-PRL}

\appendix
\begin{widetext}

\setcounter{figure}{0}

\renewcommand{\thefigure}{S\arabic{figure}}

\section{Supplemental Material}
\subsection{Bandstructure}\label{app:bs}

\begin{figure}[h]
    \centering
    \includegraphics[width=1.0\linewidth]{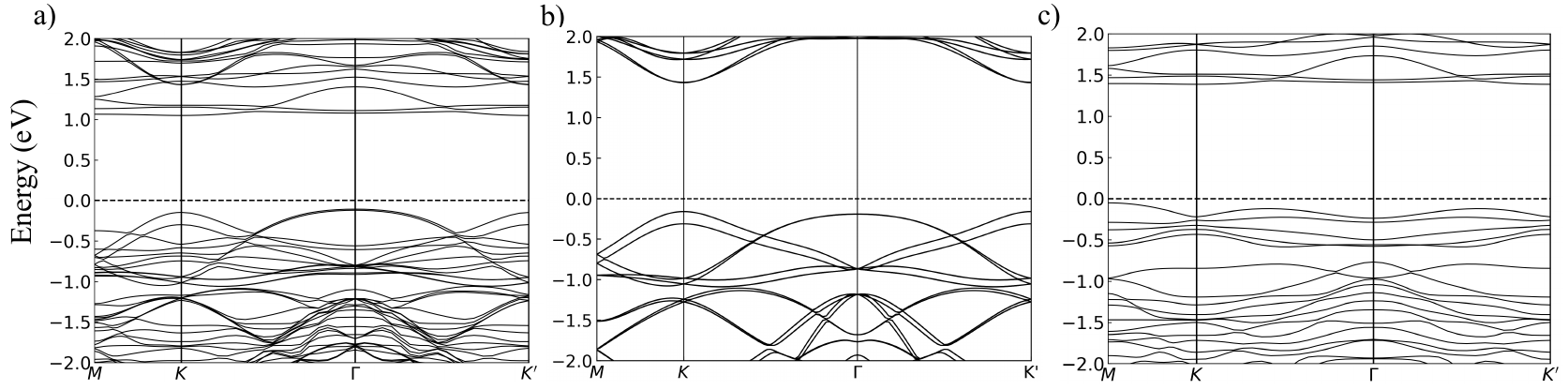}
    \caption{DFT band structure of (a) MoS$_2$/CrBr$_3$, (b) $2\times 2$ MoS$_2$, (c) monolayer CrBr$_3$ from $E_f-2$eV to $E_f+2$eV. The valence band of CrBr$_3$ is only 0.4 eV lower than the valence band of CrBr$_2$.}
    \label{fig:7}
\end{figure}

\textbf{DFT calculations. } We present the separate band structures of $2\times 2$ MoS$_2$ and CrBr$_3$ in Fig. \ref{fig:7}. The K valley is 30 meV higher than the $\Gamma$ valley in monolayer MoS$_2$. Due to the type II band alignment, the band gap of the heterobilayer is reduced to 1.2 eV. To consider the effect of Hubbard corrections, we add various values of onsite U to the present calculations with simplified (rotationally invariant) GGA + U approach \cite{dudarev1998electron}, as implemented in the Vienna Ab initio Simulation Package \cite{kresse1996efficiency}. We find the valence bands of CrBr$_3$ move slightly under Hubbard corrections, and the proximitized spin splitting is mainly determined by the valence band energy offset between CrBr$_3$ and MoS$_2$. In practice, U=1.5 eV is used in most of the recent works on CrX$_3$ (X=Cl, Br, I) family of materials \cite{wu2019physical}, and the spin splitting at $\Gamma$ valley is slightly reduced to 11 meV as shown in Fig. \ref{fig:9}, within the strong coupling regime.

\begin{figure}[h]
    \centering
    \includegraphics[width=1.0\linewidth]{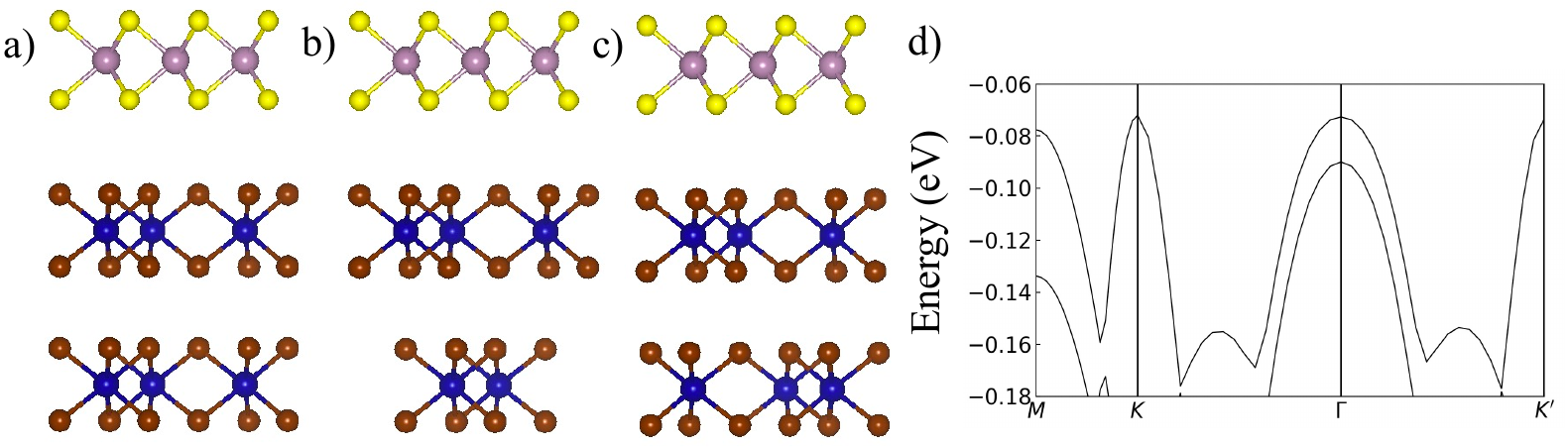}
    \caption{ (a,b,c) Lattice structures of of MoS$_2$/CrBr$_3$/CrBr$_3$ at high symmetry stacking regions, (d) Band structure of MoS$_2$/CrBr$_3$/CrBr$_3$ for stacking configuration (b).   }
    \label{fig:8}
\end{figure}

\begin{figure}
    \centering
    \includegraphics[width=0.9\linewidth]{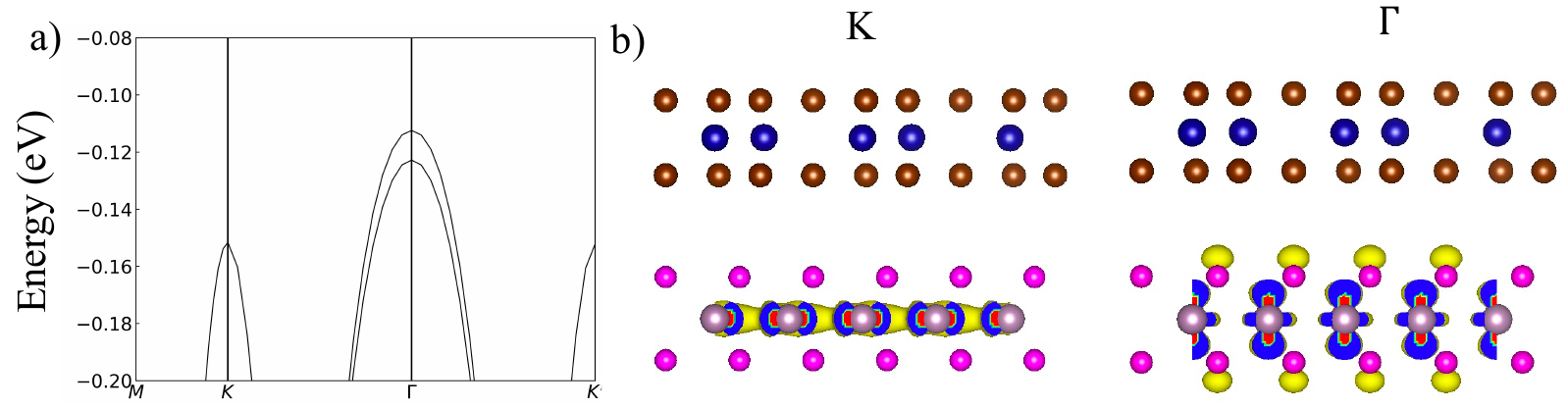}
    \caption{(a) GGA + U band structures of ferromagnetic heterobilayer CrBr$_3$ and MoS$_2$. A rotationally invariant Hubbard potential is employed with U = 1.5 eV in the GGA + U calculation. (b)Lattice structure of MoS$_2$/CrBr$_3$ with wavefunction of MoS$_2$ valence band at $\Gamma$ and $K$ momentum.}
    \label{fig:9}
\end{figure}

Unlike the $K$ valley wavefunction, which only couples to the out-of-plane magnetization, here the $\Gamma$  valley wavefunction at zero momentum is isotropic as shown in Fig. \ref{fig:9} and couples to magnetization at all directions. In DFT calculations, we check the value of proximity-induced spin splitting gap both at out-of-plane and in-plane directions and find the same value as 14 meV. We further note the small directional variations in the exchange coupling will not affect our main results. 

As shown in previous work \cite{Tong2018Nov}, Skyrmion spin textures can be realized in twisted bilayer CrBr$_3$. Therefore we further calculate the exchange splitting of MoS$_2$ on twisted bilayers CrBr$_3$. In bilayer CrBr$_3$, the bonding-antibonding splitting within valence bands pushes valence band maximum higher than monolayer CrBr$_3$, and the valence band offset between $\Gamma$ valley of MoS$_2$ and magnetic layers is threfore reduced. With new setup MoS$_2$/CrBr$_3$/CrBr$_3$, the exchange splittings are found to be around 17 meV for three high symmetry stacking configurations in Fig. \ref{fig:8}(a,b,c), slightly enhanced compared to the 14 meV splitting identified in MoS$_2$/CrBr$_3$.

\par 

\textbf{Continuum model. } The plane wave approximation for the continuum model is as follows. Let $\vec S(r) = \vec S(r+a_1)=\vec S(r+a_2)$ be a doubly periodic spin texture and let $b_1,b_2$ be the reciprocal vectors, $b_i\cdot a_j = 2\pi \delta_{ij}$.  We can write any wavefunction with momentum $k=(k_x,k_y)$ as
\begin{equation}
    \psi(r)= e^{ikr} \sum_{m_1,m_2} e^{i(m_1b_1+m_2b_2)r} \chi_{m_1m_2}
\end{equation}
where $m_1,m_2$ are integers and $\chi_{m_1m_2} = \begin{pmatrix}
    \chi^\uparrow_{m_1m_2}\\\chi^\downarrow_{m_1m_2}
    \end{pmatrix}$. The Hamiltonian can be written as 
\begin{equation}
    H_S = \frac{p^2}{2m}\mathbf{1} + J \sum_{m_1,m_2} e^{i(m_1b_1+m_2b_2)r}\vec S_{m_1m_2}\cdot \vec \sigma 
\end{equation}
where $\vec S_{m_1m_2}=\vec S^*_{-m_1,-m_2}$ is a Fourier component of $\vec S(r)$. We then obtain the matrix equation 

\begin{equation}
    \frac{(k+n_1b_1+n_2b_2)^2}{2m} \chi_{n_1n_2}+ J \sum_{m_1m_2} \vec S_{m_1m_2}\cdot \vec \sigma \chi_{n_1-m_1,n_2-m_2} = E\chi_{n_1n_2}
\end{equation}

\noindent We apply a large momentum cutoff $|n_1|,|n_2|\leq N$ to approximate the spectrum and eigenfunctions, with an analogous procedure for the Dirac case. In the Dirac case we omit spurious modes which appear on the lattice edges and whose support is measure zero as $N\to \infty$, which can be easily identified as they break the $C_6$ symmetry. 
The plane wave method was used for Fig.s \ref{fig:1}, \ref{fig:3}, and \ref{fig:4} while for Fig. \ref{fig:5} a square lattice nearest neighbor tight-binding Hamiltonian was used.

\subsection{ Small exchange coupling and anomalous Hall effect}

In the following, we connect the leading Fourier modes of point-group symmetric spin textures to the Chern number of associated electron bands. We approach this by using perturbation theory in a small exchange coupling $J$, in which case Chern numbers can be computed from the point-group symmetry eigenvalues of the states at the high-symmetry points in the Brillouin zone \cite{PhysRevB.86.115112}. These eigenvalues in turn depend only on the leading Fourier modes of the spin texture. We assume a six-fold rotational symmetry of the Hamiltonian, i.e. $[\hat C_6, H ]=0$ where $\hat C_6 = e^{\frac{-2\pi i}{6} L^z} e^{\frac{-2\pi i}{6} \frac{\sigma^z}{2}}$ (other point-groups can be analyzed similarly). The Chern number $C$ is given by

\begin{equation}
e^{2\pi iC/6} = \prod_{i\in occ.} (-1)^F\eta_i(\Gamma)\theta_i(K)\zeta_i(M)
\end{equation}
where $\eta(\Gamma)$ is the $C_6$ eigenvalue at the $\Gamma$ point, $\theta(K)$ is the $C_3$ eigenvalue at the $K$ point, and $\zeta(M)$ is the $C_2$ eigenvalue at the $M$ point. The Chern number for the first band follows from degenerate perturbation theory, and we obtain

\begin{equation}
C \bmod 6 = \begin{cases}
1 & S_0^z>0, S_{\textbf{G}}^z>0\\
2 \text{ or } 0 & S_0^z>0, S_{\textbf{G}}^z<0\\
-2 \text{ or } 0 & S_0^z<0, S_{\textbf{G}}^z>0\\
-1 & S_0^z <0, S_{\textbf{G}}^z <0\\
\end{cases}
\end{equation}

\noindent assuming $J>0$. For $J<0$ the Chern number changes sign mod 6. Here $\vec S_0, \vec S_{\textbf{G}}$ are the zeroth and first Fourier components of $\vec S(r)$, so in particular $S_0^z=\overline{m}$. Here $\textbf{G}$ is one of the primitive reciprocal lattice vectors ($\vec S_{\textbf{G}}$ doesn't depend on which one, by six-fold symmetry). For fixed $S_{\textbf{G}}^z$, it follows from the above that the first band undergoes a topological transition as the magnetization changes sign, which can be traced to the gap closing at $\Gamma$. Likewise, for fixed magnetization there is a topological transition as $S_{\textbf{G}}^z$ changes sign.

\textit{$\mathit{\Gamma}$ point.---}The two states $\ket{\Gamma,+\frac12}$ and $\ket{\Gamma,-\frac12}$ have $\hat C_6$ eigenvalues $e^{-i\pi/6}$ and $e^{i\pi/6}$ respectively. First order perturbation theory gives an energy correction 

\begin{equation}
\braket{\Gamma,s| J \vec \sigma \cdot \vec S|\Gamma,s} = 2sJ  S^z_0
\end{equation}
where $s= \pm \frac12$. Therefore 
\begin{equation}
\eta(\Gamma) = \begin{cases} e^{-i\pi/6} & JS_0^z <0\\
e^{i\pi/6} & JS_0^z>0 
\end{cases}.
\end{equation}

\textit{$\mathit{K}$ point.---}We start by defining the states

\begin{equation}
\ket{K,\nu,s} = \frac{1}{\sqrt{6}} \sum_{j=0}^5 e^{\frac{-i\pi }{6} j(2\nu+1)} \hat C_6^j \ket{K,s}\qquad \nu=0,1,\ldots, 5.
\end{equation}
 The six eigenspaces can be labelled by $\nu$:  

\begin{align*}
\hat C_6 \ket{K,\nu,s} &= e^{\frac{i\pi }{6}(2\nu+1)} \ket{K,\nu,s}
\end{align*}

The first-order effective $2\times2$ Hamiltonian in each eigenspace can be written $[H^{\text{eff.}}_\nu]_{ss'}  
= J  \braket{K,\nu,s|\vec \sigma \cdot \vec S|K,\nu,s'}$, giving

\begin{subequations}
\begin{align}
H^{\text{eff.}}_0=H^{\text{eff.}}_3 &= J \begin{pmatrix}
S_0^z +2S_{\textbf{G}}^z&0\\0& -S_0^z + S_{\textbf{G}}^z
\end{pmatrix},\\
H^{\text{eff.}}_1=H^{\text{eff.}}_4 &= J \begin{pmatrix}
S_0^z -S_{\textbf{G}}^z&0\\0& -S_0^z + S_{\textbf{G}}^z
\end{pmatrix},\\
H^{\text{eff.}}_2=H^{\text{eff.}}_5 &= J \begin{pmatrix}
S_0^z -S_{\textbf{G}}^z&0\\0& -S_0^z -2S_{\textbf{G}}^z
\end{pmatrix}.
\end{align}
\end{subequations}

Let 

\begin{equation}
\mu_K = \min[\pm J(S_0^z-S_{\textbf{G}}^z), \pm J (S_0^z+2S_{\textbf{G}}^z)].
\end{equation}

It follows that 

\begin{equation}
\theta(K) = \begin{cases}
e^{i\pi/3} & \mu_K = J(S_0^z + 2S_{\textbf{G}}^z)\\
e^{-i\pi/3} & \mu_K = -J(S_0^z + 2S_{\textbf{G}}^z)\\
-1 \text{ or } e^{-i\pi/3} & \mu_K = J(S_0^z-S_{\textbf{G}}^z)\\
e^{i\pi/3} \text{ or } -1 & \mu_K= -J(S_0^z-S_{\textbf{G}}^z)
\end{cases}.
\end{equation}
The latter two cases require 2nd order perturbation theory, and thus higher Fourier harmonics, to resolve. 

\textit{$\mathit{M}$ point.---}Define the states 

\begin{equation}
\ket{M, \nu, s} = \frac{1}{\sqrt{6}} \sum_{j=0}^5 e^{\frac{-i\pi}{6}j(2\nu+1)} \hat C_6^j \ket{M,s}\qquad \nu=0,1,\ldots,5.
\end{equation}

Once again the six eigenspaces can be labelled by $\nu$:

\begin{equation}
\hat C_6 \ket{M,\nu,s} = e^{\frac{i\pi}{6}(2\nu+1)} \ket{M,\nu,s}.
\end{equation}

The first-order effective Hamiltonian in each eigenspace is given by $[H^{\text{eff.}}_\nu]_{ss'}  =\braket{M,\nu,s|V|M,\nu,s'}$, yielding

\begin{subequations}
\begin{align}
H^{\text{eff.}}_0=H^{\text{eff.}}_2= H^{\text{eff.}}_4&= J \begin{pmatrix}
S_0^z +S_{\textbf{G}}^z&0\\0& -S_0^z + S_{\textbf{G}}^z
\end{pmatrix},\\
H^{\text{eff.}}_1=H^{\text{eff.}}_3=H^{\text{eff.}}_5 &= J \begin{pmatrix}
S_0^z -S_{\textbf{G}}^z&0\\0& -S_0^z - S_{\textbf{G}}^z
\end{pmatrix}.
\end{align}
\end{subequations}
Let
\begin{equation}
\mu_M = \min[\pm J(S_0^z+S_{\textbf{G}}^z),\pm J (S_0^z-S_{\textbf{G}}^z)].
\end{equation}
It follows that 

\begin{equation}
\zeta(M) = \begin{cases}
i & \mu_M =  J(\pm S_0^z + S_{\textbf{G}}^z) \\
-i & \mu_M = J(\pm S_0^z - S_{\textbf{G}}^z)
\end{cases}.
\end{equation}

\section{Berry curvature}

The profile of Berry curvature in $k$-space is essential for studies of strongly correlated physics in Chern bands. For example, a fractional Chern insulator is more strongly favored when the Berry curvature is flat (i.e. closer to the ideal case of a Landau level)\cite{Parameswaran2013Nov}. Motivated by these concerns, we plot the Berry curvature for some textures explored in the main text in Fig. \ref{fig:SMberry}. We observe that at the ``magic" magnetization $\bar m =0.22$ where the lowest band becomes very flat, the Berry curvature becomes significantly more evenly distributed (than at $\bar m =0.1$, for instance). The question of whether a fractional Chern insulator can be realized in this system is an interesting direction for future studies.

\begin{figure}[b!]
    \centering  \includegraphics[width=0.6\linewidth]{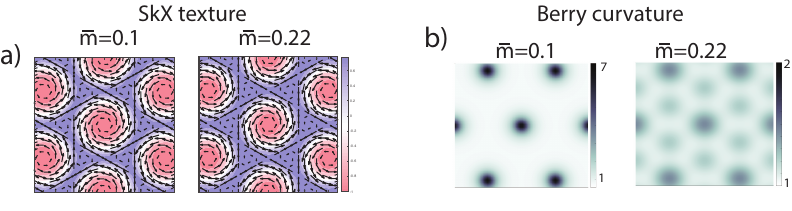}
    \caption{(a) Skyrmion crystal textures reproduced from Fig. \ref{fig:2}a. (b) Rescaled Berry curvature $|b_1|^2\Omega(\bk)/2\pi$ corresponding to the lowest bands in in Fig. \ref{fig:2}b plotted in a square of size $2|b_1|$. Berry curvature becomes more uniform near the flat band at $\overline{m}=0.22$.
}
    \label{fig:SMberry}
\end{figure}

\end{widetext}

\end{document}